\documentclass{article}

\usepackage[nonatbib,final]{nips_2017}

\usepackage[]{graphicx}
\usepackage{subcaption}

\usepackage[utf8]{inputenc} 
\usepackage[T1]{fontenc}    
\usepackage{hyperref}       
\usepackage{url}            
\usepackage{booktabs}       
\usepackage{amsfonts}       
\usepackage{nicefrac}       
\usepackage{microtype}      
\usepackage{varwidth}

\title{ShortScience.org - Reproducing Intuition}

\author{
Joseph Paul Cohen\\
Institute for Reproducible Research\\
and\\
Montreal Institute for Learning Algorithms\\
Universit\'{e} of Montr\'{e}al\\
\texttt{cohenjos@iro.umontreal.ca} \\
\And
Henry Z. Lo \\
Institute for Reproducible Research\\
\texttt{henryzlo@cs.umb.edu}
}

\begin{document}

\maketitle

\begin{abstract}
We present ShortScience.org, a platform for post-publication discussion of research papers.  On ShortScience.org, the research community can read and write summaries of papers in order to increase accessible and reproducibility. Summaries contain the perspective and insight of other readers, why they liked or disliked it, and their attempt to demystify complicated sections. ShortScience.org has over 600 paper summaries, all of which are searchable and organized by paper, conference, and year.  Many regular contributors are expert machine learning researchers.  We present statistics from the last year of operation, user demographics, and responses from a usage survey.  Results indicate that ShortScience benefits students most, by providing short, understandable summaries reflecting expert opinions.  

\end{abstract}

\section{Overview}

\texttt{ShortScience.org} is a platform for post-publication discussion of research papers.  Users can write summaries for research papers on the site.  Interested readers can read these summaries to get multiple perspectives on the given paper, in addition to the author's, thus gaining better understanding.  Many regular contributors are expert machine learning researchers, whose descriptions make papers, and by extension the field of research, more accessible for all.

Papers can be hard to understand, for a variety of reasons:

\begin{itemize}
\item Different communities have different nomenclature to describe the same concepts
\item There is a lot of jargon in papers, often making vanilla ideas sound novel
\item Some ideas are very complex and could use multiple perspectives to get a more complete understanding
\item Some parts of ideas may be obscure so that flaws in papers cannot be found
\item Authors are encouraged to make the work seem as significant and important as possible for it to be accepted
\item Some readers do not have access to papers directly and rely on second hand knowledge
\end{itemize}








Asking multiple domain experts to explain is an excellent way to understand a piece of research.  However, not everyone has access to an expert, let alone multiple.  ShortScience.org provides a platform for experts and non-experts alike to share notes on papers.  These notes are available to all, providing a variety of explanations to help everyone better understand. 






\begin{figure}

    \centering
    \fbox{
    \includegraphics[width=0.9\textwidth]{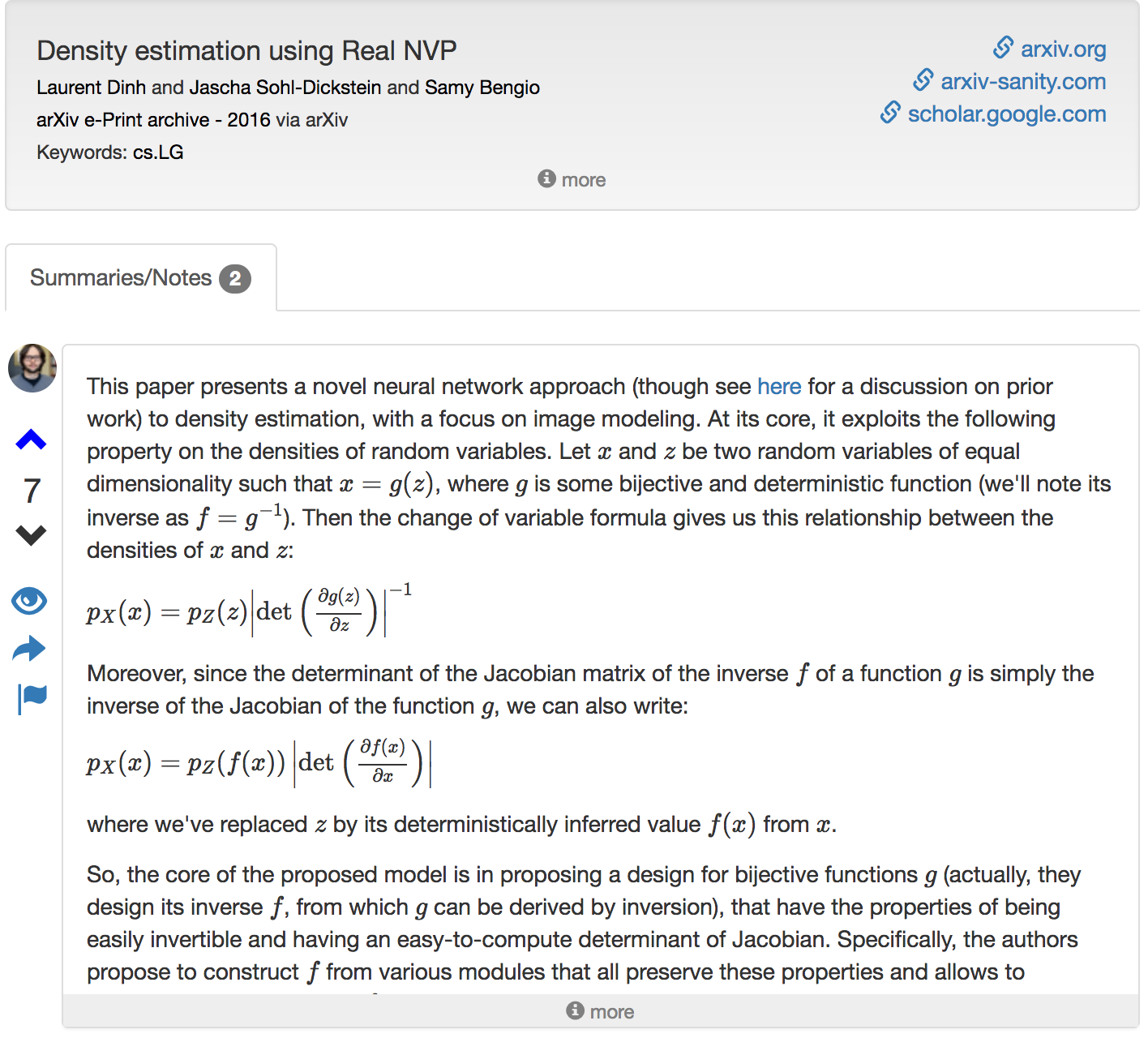}%
    }
    
    \caption{Example summary available on \texttt{ShortScience.org}. Each paper has information such as venue, abstract, and useful links followed by a summary box.  The summary contains author, votes, view-source and the summary text, which can be formatted in Markdown and contain {\LaTeX}  math, images, and videos.}
    \label{fig:example}
\end{figure}

\newpage
\section{Approach}

The \texttt{ShortScience.org} platform provides three main features:

\begin{itemize}
\item Post summaries/notes on papers (public, private, or anonymous)
\item Comment on summaries/notes
\item Search, browse by venues, and follow users
\end{itemize}

Summaries can be written for any paper in three main databases, which includes anything with a DOI, on ar$\large\chi$iv, or on Bibsonomy \cite{Hotho2006}. These summaries can be voted on by each user using a simple up or down metric. Each summary can be set as private which is useful for personal organization of papers. 

\texttt{ShortScience.org} is run and managed by the Institute for Reproducible Research (IRR), a U.S. Non-Profit organization. The IRR also manages the project \texttt{academictorrents.com} which is a system that facilitates the movement of large datasets for research \cite{Lo2016,Cohen2014}.

\begin{figure}
    \centering

    \hspace{20pt}
    \begin{subfigure}[t]{0.33\textwidth}
    \centering\captionsetup{width=0.95\linewidth, singlelinecheck=false}
    \includegraphics[width=1\textwidth]{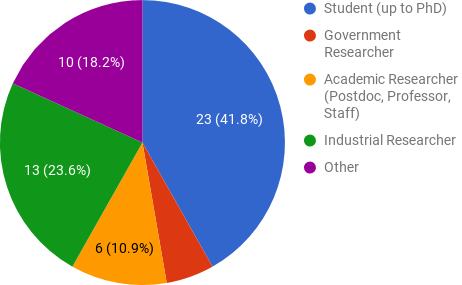}%
    \caption*{Current Employment}
    \label{fig:employment}
    \end{subfigure}%
    \begin{subfigure}[t]{0.33\textwidth}
    \centering\captionsetup{width=.6\linewidth, singlelinecheck=false}
    \includegraphics[width=1\textwidth]{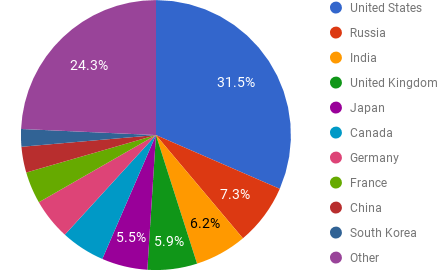}%
    \caption*{Geographic}
    \label{fig:geo}
    \end{subfigure}%
    \begin{subfigure}[t]{0.33\textwidth}
    \centering\captionsetup{width=.4\linewidth, singlelinecheck=false}
    \includegraphics[width=1\textwidth]{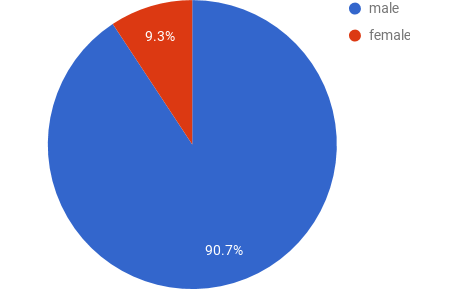}%
    \caption*{Gender}
    \label{fig:gender}
    \end{subfigure}%
    
    \vspace{10pt}
    \begin{subfigure}[t]{0.45\textwidth}
    \includegraphics[width=1\textwidth]{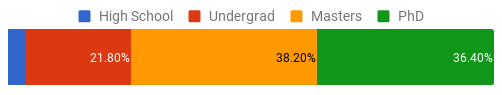}%
    \caption*{Academic degree (current or highest obtained)}
    \label{fig:degree}
    \end{subfigure}%
    \hspace{10pt}
    \begin{subfigure}[t]{0.45\textwidth}
    \includegraphics[width=1\textwidth]{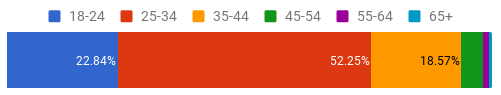}%
    \caption*{Age}
    \label{fig:age}
    \end{subfigure}%
    
    \vspace{10pt}
    \begin{subfigure}[t]{1.0\textwidth}
    \centering\captionsetup{}
    \includegraphics[width=1\textwidth]{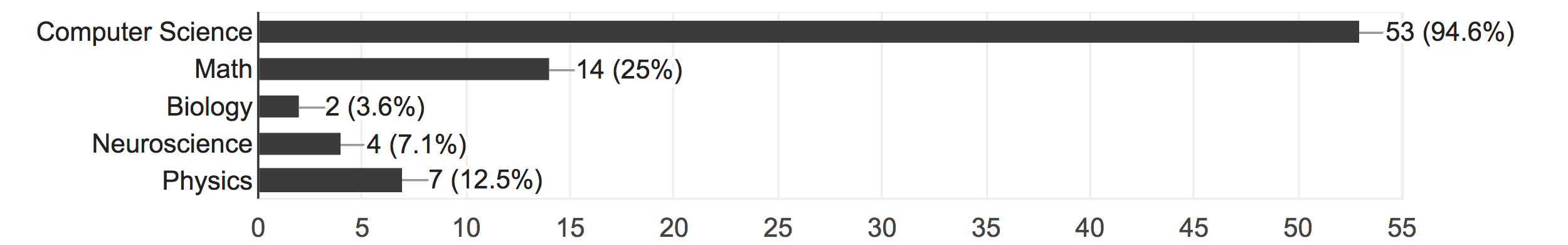}%
    \caption*{What field(s) do you study?}
    \label{fig:fields}
    \end{subfigure}%

    \caption{User demographics, collected using a survey and website statistics}
    \label{fig:demographics}
\end{figure}

\section{Community Impact}

Over the last year of the site's operation, \texttt{ShortScience.org} has received 34,938 unique users to the 626 public and 83 private summaries. These users visited the site 118,874 times and spent an average duration of 1.41 minutes per visit. These users come from all over the world, are mainly focused in Computer Science, typically enrolled in Masters or PhD programs, and younger than 30. More detailed demographics are shown in Figure \ref{fig:demographics}. Based on a sample of 55 users, we found:

\begin{itemize}
\item 60\% of users read 5 or more summaries
\item 87\% of users found reading these summaries useful in understanding papers
\item \textbf{82\% of users read summaries for papers that they would not have otherwise read}
\end{itemize}

These usage statistics suggest that summaries are helpful for both readers, in terms of understanding, and for authors in terms of readers reached.



\subsection{Gender}

Users were only 9.3\% are female. Because the primary content on the site is Machine Learning related, this may reflect a trend in Machine Learning that differs from Computer Science as a whole. The National Science Board's Science and Engineering Indicators report \cite{NationalScienceBoard2016} states 25.3\% (671,000/2,647,000) are employed as computer and mathematical scientists in 2016. Supporting this number, the Survey of Earned Doctorates \cite{NationalScienceFoundation2015} reports 24\% (943/3,825) earned a PhD in mathematics and computer sciences in 2015. These numbers indicate a bias in Machine Learning.

\subsection{Reproducibility}
\label{sec:rep}

We define reproducibility as recreating the intuition the author tried to describe in their paper and as recreating the experiments in order to verify results. Recreating an experiment alone will not guarantee the intuition can be passed on to the reader, however recreating the intuition directly can enable a research to implement their own solution to verify results. 

We assess intuition reproducibility explicitly with user reported success in Figure \ref{fig:sentiment}. In our survey we found 87\% of users were able to use the platform to understand a research paper. While the majority of users did not try to directly reproduce research using the site, 10.9\% (6/55 users surveyed) did and were successful while 5.5\% (3/55) reported the platform not helping them and 83.6\% (46/55) did not try to reproduce results.

\begin{figure}

    \centering
    \begin{subfigure}[t]{0.33\textwidth}
    \centering\captionsetup{width=.9\linewidth}
    \includegraphics[width=1\textwidth]{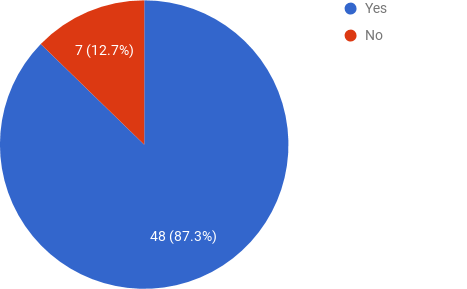}%
    \caption{Has ShortScience.org helped you in understanding research?}
    \label{fig:helped-understand}
    \end{subfigure}%
    \begin{subfigure}[t]{0.33\textwidth}
    \centering\captionsetup{width=.9\linewidth}
    \includegraphics[width=1\textwidth]{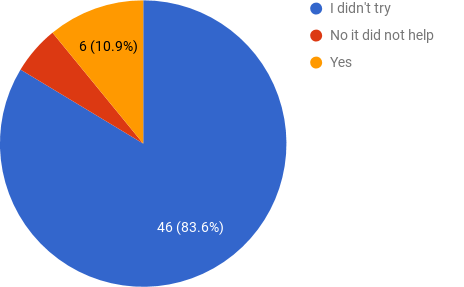}%
    \caption{Has ShortScience.org helped you reproduce the results of a published work?}
    \label{fig:helped-reproduce}
    \end{subfigure}%
    \begin{subfigure}[t]{0.33\textwidth}
    \centering\captionsetup{width=.9\linewidth}
    \includegraphics[width=1\textwidth]{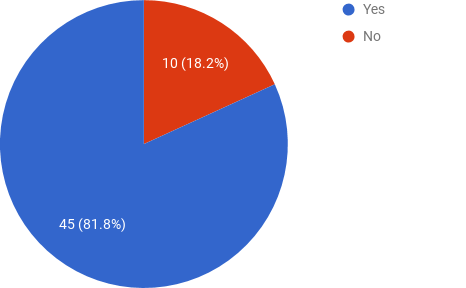}%
    \caption{Have you read a summary for a paper you would otherwise have not read?}
    \label{fig:otherwise}
    \end{subfigure}%

    \vspace{20pt}
    \begin{subfigure}[t]{0.6\textwidth}
    \includegraphics[width=1\textwidth]{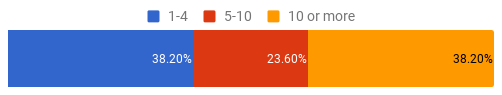}%
    \caption{How many summaries have you read on ShortScience.org?}
    \label{fig:how-many}
    \end{subfigure}%

    \caption{Questions on usage}
    \label{fig:my_label}
\end{figure}

\subsection{Usefulness}
\begin{figure}[h]
    \centering
    
    \begin{subfigure}[t]{0.5\textwidth}
    \centering\captionsetup{width=.8\linewidth}
    \includegraphics[width=1\textwidth]{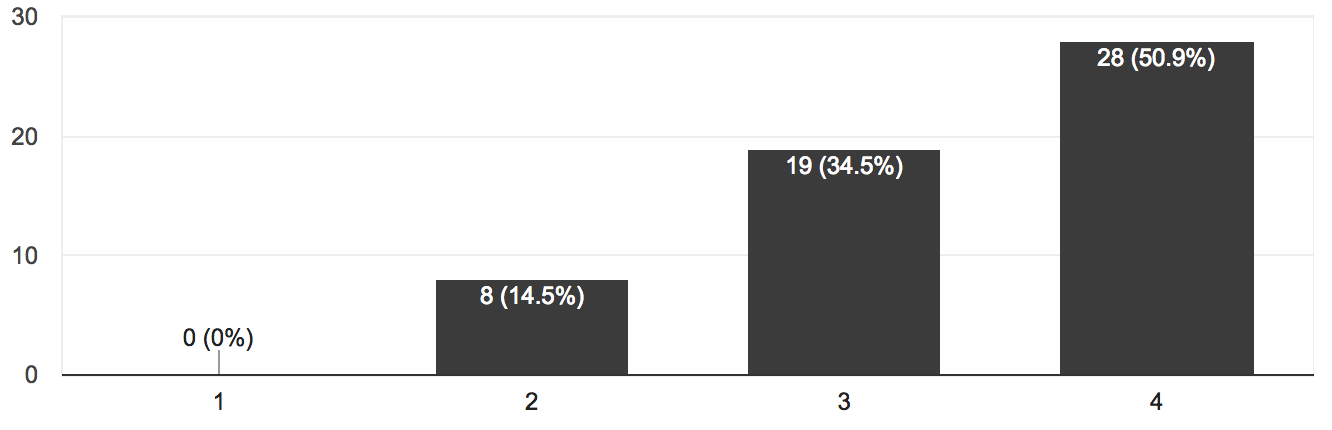}%
    \caption{How useful is ShortScience.org for you?}
    \label{fig:useful}
    \end{subfigure}%
    \begin{subfigure}[t]{0.5\textwidth}
    \centering\captionsetup{width=.8\linewidth}
    \includegraphics[width=1\textwidth]{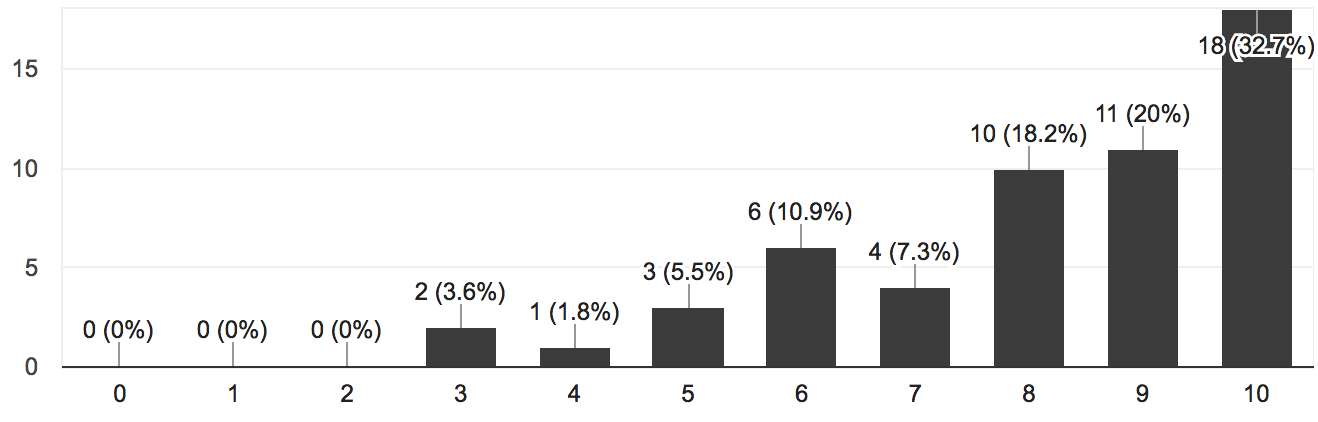}%
    \caption{How likely are you to recommend ShortScience.org to a friend or colleague?}
    \label{fig:nps}
    \end{subfigure}%

    \caption{General sentiment towards the site}
    \label{fig:sentiment}
\end{figure}

Responses from the survey (\ref{fig:useful}) indicate that the project is perceived to be useful. A more detailed version of this poll is shown in Figure \ref{fig:nps} which allows us to use the Net Promoter Score (NPS) evaluation \cite{Keiningham2008}.  NPS asks the question "How likely are you to recommend \texttt{ShortScience.org} to a friend or colleague?" and present 11 choices between 0 and 10. From the responses, the NPS is calculated as $\frac{\textnormal{\# promoters} - \textnormal{\# detractors}}{\textnormal{\# total respondents}}$ where promoters are those who responded $9-10$ and detractors responded between $0-6$. The European variant accounts for respondents giving lower scores even though they are satisfied and alters these numbers to $8-10$ and $0-5$. We observe a score of 31 using the U.S. scale and 60 using the European variant. The range of possible scores are between $-100$ and $+100$, so the observed scores are fairly good. 
\vspace{20pt}

\section{Conclusion}
Here we presented \texttt{ShortScience.org}, which aims to make research more accessible by making the ideas more understandable.  After one year of operation the site has made impact, as measured by survey results. 82\% of users read summaries for papers that they would not have otherwise read. The project has also helped 87\% of users understand the research papers they are reading and 10.9\% directly reproduce results of a paper. The project has impact on the machine learning community and is expected to have more in the future.

\bibliographystyle{ieee}
\bibliography{publishing}

\end{document}